\documentclass[%
 reprint,
 twocolumn,
 superscriptaddress,
 amsmath,amssymb,
 aps,
 prl
]{revtex4-1}

\usepackage{graphicx}
\usepackage{dcolumn}
\usepackage{bm}
\usepackage{hyperref}

\usepackage{physics}
\usepackage{verbatim}
\usepackage{xcolor, soul}

\usepackage[outdir=./]{epstopdf}
\usepackage[export]{adjustbox}

\begin{document}

\title{Exact Quantum Dynamics in Structured Environments}

\author{Dominic Gribben}
 \affiliation{SUPA, School of Physics and Astronomy, University of St Andrews,
St Andrews, KY16 9SS, United Kingdom}
\author{Aidan Strathearn}
 \affiliation{SUPA, School of Physics and Astronomy, University of St Andrews,
St Andrews, KY16 9SS, United Kingdom}
 \affiliation{School of Mathematics and Physics, The University of Queensland, St Lucia,
Queensland 4072, Australia}
\author{Jake Iles-Smith}
 \affiliation{Department of Physics and Astronomy,  University of Sheffield, Sheffield,  S3  7RH,  United  Kingdom}
\author{Dainius~Kilda}
 \affiliation{SUPA, School of Physics and Astronomy, University of St Andrews,
St Andrews, KY16 9SS, United Kingdom}
\author{Ahsan Nazir}
 \affiliation{School  of  Physics  and  Astronomy,  The  University  of  Manchester,  Oxford  Road,  Manchester  M13  9PL,  United Kingdom}
\author{Brendon W. Lovett}
 \affiliation{SUPA, School of Physics and Astronomy, University of St Andrews,
St Andrews, KY16 9SS, United Kingdom}
\author{Peter Kirton}
\affiliation{Vienna Center for Quantum Science and Technology, Atominstitut, TU Wien, 1040 Vienna, Austria}

\date{\today}

\begin{abstract}
The dynamics of a wide range of technologically important quantum systems are dominated by their interaction with just a few environmental modes. Such highly structured environments give rise to long-lived bath correlations that induce complex dynamics which are very difficult to simulate.
These difficulties are further aggravated when spatial correlations between different parts of the system are important.
By modeling the dynamics of a pair of two-level quantum systems in a common, structured, environment we show that a recently developed numerical approach, the time-evolving matrix product operator, is capable of accurate simulation under exactly these conditions. 
We find that tuning the separation to match the wavelength of the dominant environmental modes can drastically modify the system dynamics. 
To further explore this behavior, we show that the full dynamics of the bath can be calculated directly from those of the system, thus allowing us to develop intuition for the complex system dynamics observed.
\end{abstract}

\maketitle

When a quantum system interacts with a structured environment in which a narrow band of modes dominate, the resulting dynamics can be very complex and difficult to simulate accurately. Such environments can retain a memory of their interactions with the system on a timescale comparable to that on which the state of the system changes, and in the face of such memory effects standard open system techniques can fail.
However, reliable simulation of such environments is vital in order to understand the behavior of an ever-increasing number of  experimental platforms and quantum devices.
For example, in photosynthetic systems the observation of quantum coherence comes from strong electronic and vibrational coupling in biomolecules~\cite{turner11}, where the interplay of these degrees of freedom has been shown to enhance energy transport~\cite{thorwart2009enhanced, chin2013role, rey2013exploiting, mohseni2014energy}.
Similar physics can also be found in e.g.\ trapped ions~\cite{maier2019environment} and molecular junctions~\cite{segal2016vibrational}.
Further, bath memory plays a key role in the function of micromechanical resonators, where adjustment of the environmental noise  spectrum is possible~\cite{groblacher2015observation} and photonic crystals, where band gaps in the spectral density give rise to localized modes and dissipationless oscillations~\cite{zhang2012general, Goban14, Sundaresan19}. 
The effect of structured environments has also been explored with superconducting qubits subject to different kinds of noise~\cite{potovcnik2018studying}.

Memory effects can be even more pronounced when the system coupled to the structured environment described above is extended.
For example if it consists of several spatially separated sites. 
The dynamics then become sensitive to groups of environmental modes with a narrower bandwidth, and so a longer correlation time. This sensitivity was recently observed in an experiment on a `giant' superconducting atom~\cite{andersson2018nonexponential}. 
Spatial correlation effects have also been studied in the context of quantum transport~\cite{nalbach2010quantum,nazir2009correlation,mccutcheon2011coherent,abramavicius2011exciton,sarovar2011environmental}, and
spatially correlated noise in quantum registers has been shown to affect quantum error propagation~\cite{aharonov2006fault}.

In order to model environmental memory effects
, we must go beyond the standard Born-Markov approximations~\cite{breuer2002theory} which lead to a simple time-local master equation for the system density matrix. 
The development of techniques to simulate these non-Markovian dynamics has therefore been the subject of much relatively recent theoretical effort~\cite{de2017dynamics}. 
These techniques broadly fall into two categories:
First, there are approximate methods which change the boundary between the system and environment such that the Born-Markov approximations are valid for the new system~\cite{silbey1984variational,harris1985variational, jang2008theory, iles2014environmental,iles2015excitation,iles2016energy};
second, there are numerically exact methods which utilize some particular structure of the bath Hamiltonian to provide exact dynamics~\cite{tanimura1989time, makri1995tensor,makri1995tensor-2, prior10, strathearn2017efficient,mascherpa2019}. 
These approaches usually work best for models where the system Hilbert space is quite small, or has a specific form of spectral density. 
However, it can often be difficult to know \textit{a priori} the range of validity of some of  these approaches, and so  accurate and efficient benchmarking procedures are essential.

In this Letter, we demonstrate that a recently developed exact approach, the Time-Evolving Matrix Product Operator (TEMPO)~\cite{strathearn2017efficient} can be used to efficiently find non-local system dynamics for highly structured environments. To do this, we study a model consisting of a pair of spatially separated two-level subsystems interacting with a common environment, in which a narrow band of modes dominate the interaction. 
The complex environmental structure and spatial correlations give rise to highly non-Markovian dynamics, which we are able to explore using TEMPO.
In addition, we find that by tuning the separation between the subsystems, it is possible to control their interaction with the environment.

An advantage of techniques that include some environmental modes in the system description is that they enable the extraction of some of the bath's dynamics, which can in turn lead to insight into the system's behavior through understanding system-environment correlations~\cite{iles2014environmental,chin2013role}. However, we will show here that exact environmental dynamics can be easily and efficiently extracted from TEMPO, leading us to a full picture of the complex quantum dynamics of this model.

The Hamiltonian for our model is given by:
\begin{multline}\label{eq:fullham}
H =  \sum_{i} \epsilon_i \dyad{X_i}{X_i} + \frac{\Omega}{2} \left( \dyad{X_1}{X_2}+\dyad{X_2}{X_1} \right)\\
+ \sum_{i} \sigma_{z,i} \sum_k (g_{i,k} a_{k} + g_{i,k}^* a_{k}^{\dagger}) + \sum_{k} \omega_{k} a_{k}^{\dagger} a_{k}.
\end{multline}
The two-level system (TLS) at site $i$ and position $r_i$ has energy splitting $\epsilon_i$ and an excited (ground) state denoted $\ket{X_i}(\ket{0_i})$. The pair of TLSs form a dimer with a coherent coupling of strength $\Omega$, and 
the environment consists of bosons confined to one dimension: $a_{k}^{\dagger}$ creates an excitation in the mode with wave vector $k$. We also introduce the Pauli operators $\sigma_{z,i} = \dyad{X_i}{X_i}-\dyad{0_i}{0_i}$. The coupling $g_{i,k} =g_k \exp(-i k r_i)$ consists of a position independent part $g_k$ and a phase that depends on the site $i$. 
This kind of model could underpin a wide range of physical systems, for example
biological or molecular systems undergoing energy transport and interacting with vibrational modes~\cite{rey2013exploiting}, superconducting qubits in microwave resonators~\cite{zheng13}, or quantum dots interacting with a micromechanical resonator~\cite{yeo13}.

The bath can be completely characterized by its spectral density
\begin{align}\label{eq:spectral}
J(\omega) = \sum_k |g_k|^2 \delta (\omega-\omega_k),
\end{align}
from which we can calculate its autocorrelation function in thermal equilibrium at temperature $T$:
\begin{align}\label{eq:corr}
C(t) = \int_0^{\infty} d\omega J(\omega)\left(\coth\left(\frac{\omega}{2T}\right)\cos(\omega t) - i \sin(\omega t)\right).
\end{align}
If this function decays slowly compared to the system timescales then the bath memory is important.

We now consider a bath in which a narrow band of modes dominate the interaction with the system,
giving rise to the spectral density~\cite{mukamel1995principles}
\begin{align}\label{eq:underdamped}
J_0(\omega) = \frac{\alpha \Gamma \omega_0^2 \omega}{(\omega_0^2-\omega^2)^2+\Gamma^2 \omega^2}.
\end{align}
Here $\alpha$ gives the coupling strength, $\omega_0$ is the frequency characterizing the dominant mode and $\Gamma$ provides a measure of the width of the dominant band.

The Hamiltonian, Eq.~\eqref{eq:fullham}, does not couple states with a different total number of system excitations and so we restrict ourselves to the single excitation subspace, the only one where non-trivial dynamics are observed. Within this subspace it is possible to map the problem onto that of a single spin coupled to a bosonic environment~\cite{stace05}.
The mapped Hamiltonian is then a spin-boson Hamiltonian
\begin{align}\label{eq:spinbosham}
H = \frac{\Omega}{2} \sigma_x + \frac{\epsilon}{2} \sigma_z + \sigma_z \sum_k  (\tilde{g}_k a_k+\tilde{g}_k^* a_k^{\dagger})+\sum_k \omega_k a_k^{\dagger} a_k,
\end{align}
where $\epsilon=\epsilon_2-\epsilon_1$ and $\tilde{g}_k=2 i g_{k}\sin(k(r_1-r_2)/2)$. 
We also define new Pauli operators: $\sigma_z=(\sigma_{z,2}-\sigma_{z,1})/2$ 
and $\sigma_x=\dyad{X_2}{X_1}+\dyad{X_1}{X_2}$. This modification to the coupling constants results in a renormalization of the spectral density 
\begin{align}\label{eq:commonenviron}
J (\omega) =2 J_0(\omega) (1-\cos (\omega R))
\end{align}
where $R=|r_1-r_2|$ is the dimer separation and we have assumed a linear dispersion $\omega(k)=c|k|$ with $c=1$. The cosine term arises from the phase factors in the original coupling terms.

\begin{figure}
\includegraphics[width=\linewidth]{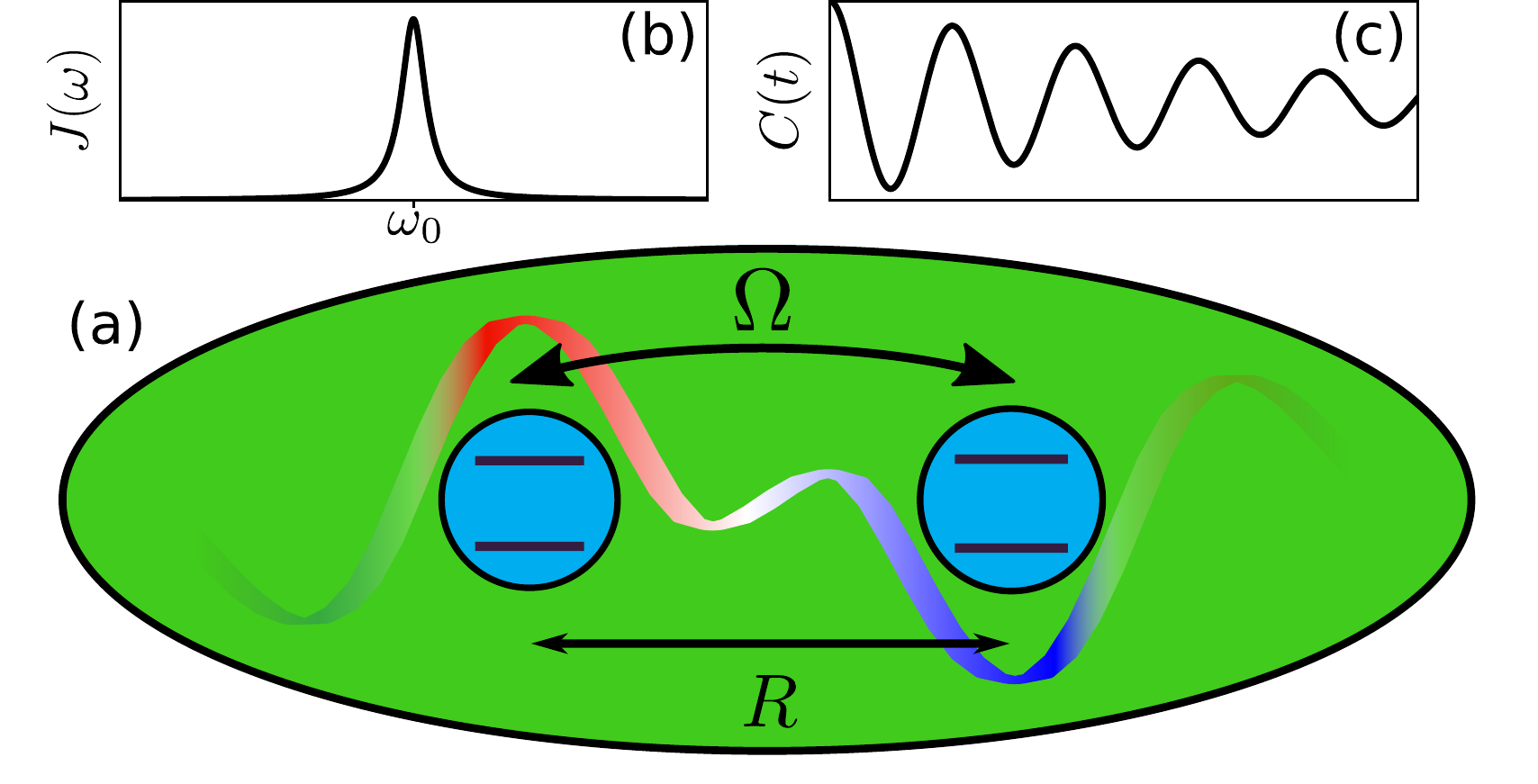}
\caption{(a) Cartoon of the dimer system we consider accompanied with schematic plots of (b) the under-damped spectral density $J(\omega)$ and (c) the corresponding bath auto-correlation function $C(t)$. 
}
\end{figure}

We begin by studying a simplified case of a single spin described by Eq.~\eqref{eq:spinbosham} interacting with a bath with the {\it unmodified} spectral density, Eq.~\eqref{eq:underdamped}.
It is straightforward to simulate this case with the reaction-coordinate (RC) mapping, which has been rigorously benchmarked for similar problems~\cite{iles2014environmental,iles2015excitation,iles2016energy}, allowing us to verify that TEMPO is able to accurately simulate dynamics in structured environments.
This mapping takes a single collective mode of the environment (the RC) into the system definition; then the rest of the bath, which is assumed to be Markovian, couples to the now-augmented system. More details can be found in Ref.~\footnote{Supplementary material which cites Refs.~\cite{garg1985effect, leggett1984quantum} and contains details of the RC mapping, numerical methods used and derivation of the expression for the bath displacement.}.
The RC mapping can be performed analytically for the spectral density in Eq.~\eqref{eq:underdamped}~\cite{iles2014environmental}, and is expected to give accurate results in the cases we present here.  
To simulate more complex environments would require more RCs in the augmented system Hilbert space, whose dimension grows exponentially with the number of included RCs.

An alternative, numerically exact way of simulating the system dynamics arising from Eq.~\eqref{eq:spinbosham} is to use a Feynman sum-over-histories approach, using an influence functional to capture environmental memory effects. 
By writing this path integral representation as a sum over discrete time steps, we can calculate the system quantum dynamics by contracting a tensor network: this is the TEMPO approach~\cite{strathearn2017efficient}.  
The memory effects are then encoded in a matrix product state which contains information about the history of the system.
This is propagated forward in time by successive contraction with a matrix product operator.
By performing singular value decompositions and truncation after each timestep~\cite{orus2014practical,schollwock2011density} we can capture bath memory times orders-of-magnitude beyond those possible with previous path integral approaches~\cite{makri1995tensor, makri1995tensor-2}. 
The similarities between TEMPO and the process tensor~\cite{Pollock2018} have recently been explored~\cite{Jorgensen2019exploiting}, and TEMPO has also been used to model dynamics in optomechanical systems~\cite{minoguchi2019environment}. 
Further details on TEMPO can be found in Ref.~\cite{Note1}.

\begin{figure}
\includegraphics[width=\linewidth]{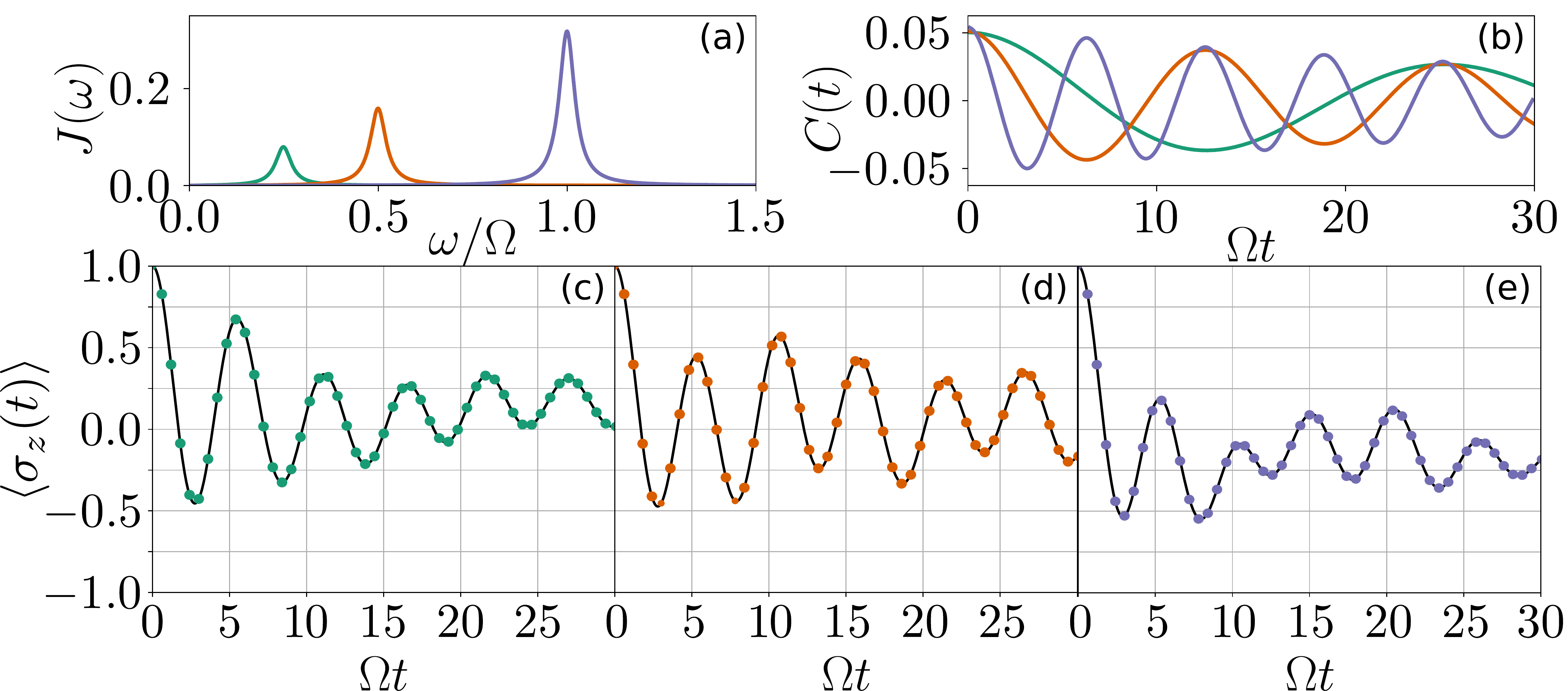}
\caption{\label{fig:underdamped} (a) Underdamped spectral density, given by Eq.~\eqref{eq:underdamped}, with the different frequencies used. 
(b) Corresponding correlation functions. 
(c)-(e) Dynamics calculated using TEMPO (dots) and RC (solid lines) for $\omega_0 = \Omega/4, \Omega/2, \Omega$.
The other parameters are $\pi \alpha=0.05\Omega$, $\Gamma = 0.05\Omega$, $T = \Omega$, $\epsilon=0.5\Omega$.}
\end{figure}

In Fig.~\ref{fig:underdamped} we compare dynamics generated by these two approaches for the simplified model of the single spin interacting with the structured spectral density, for three different values of $\omega_0$. 
We find excellent agreement between the two algorithms for all sets of parameters.
The RC approach is accurate for these parameters and this form of spectral density, and so able to capture the complex dimer dynamics. In this simple case RC requires significantly reduced numerical resources compared to TEMPO,
though we find that both techniques take longer to converge at lower mode frequencies and higher temperatures. 
For the RC algorithm the reason for this is clear: the occupation of the mode included in the system increases in these regimes and so a larger system Hilbert space is needed for convergence.
For TEMPO we find that the number of singular values that need to be retained for convergence increases -- i.e.\ more system paths gain a significant amplitude, as expected when the environment becomes more occupied.
These results show both that the RC approach is able to accurately simulate dynamics for this model~\cite{iles2016energy} and that TEMPO is able to deal with such structured spectral densities.

We now turn to analyzing the full model including the spatial modifications to the spectral density (i.e. using Eq.~\ref{eq:commonenviron}). Previous studies have shown that spatial correlations in the bath can have a significant effect on population transfer between localized systems~\cite{nalbach2010quantum,nazir2009correlation,mccutcheon2011coherent, strathearn2017efficient}.
However, these studies focused on the case where the original spectral density $J_0(\omega)$ is much broader than that given in Eq.~\eqref{eq:spectral} -- i.e. they do not focus on systems where just a narrow band of modes dominate the system interaction -- and so the modification due to the spatial structure of the system is not as pronounced. 

Since we are now using the general mapped spectral density in Eq.~\eqref{eq:commonenviron}, treating the dynamics using the RC formalism becomes much more difficult: to accurately capture the dynamics it would be necessary to include more modes in the system than is computationally feasible.
This means that for the remainder of this Letter all results are obtained using TEMPO.

\begin{figure}[t]
\includegraphics[width=\linewidth]{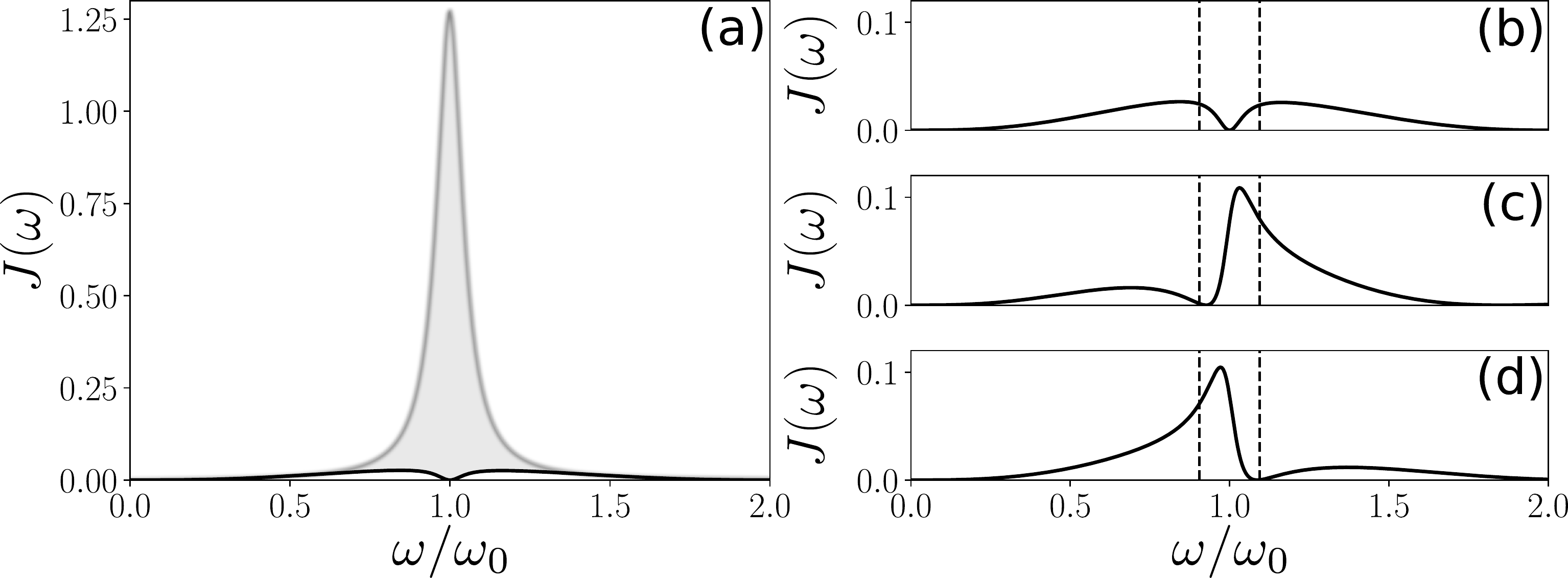}
\caption{\label{fig:CommonEnvironSpec} Effect of the mapping, Eq.~\eqref{eq:commonenviron}, on the spectral density. 
(a) The unmapped spectral density (grey shaded) and an example of the mapped case (solid black).
(b)-(d) Mapped spectral densities for the three cases considered in the main text from top to bottom: $R = 2\pi/\omega_0$, $2\pi/1.1\omega_0$ and $2\pi/0.9\omega_0$. The dashed black lines depict the frequencies described in the main text.}
\end{figure}

\begin{figure}[t!]
\includegraphics[width=\linewidth]{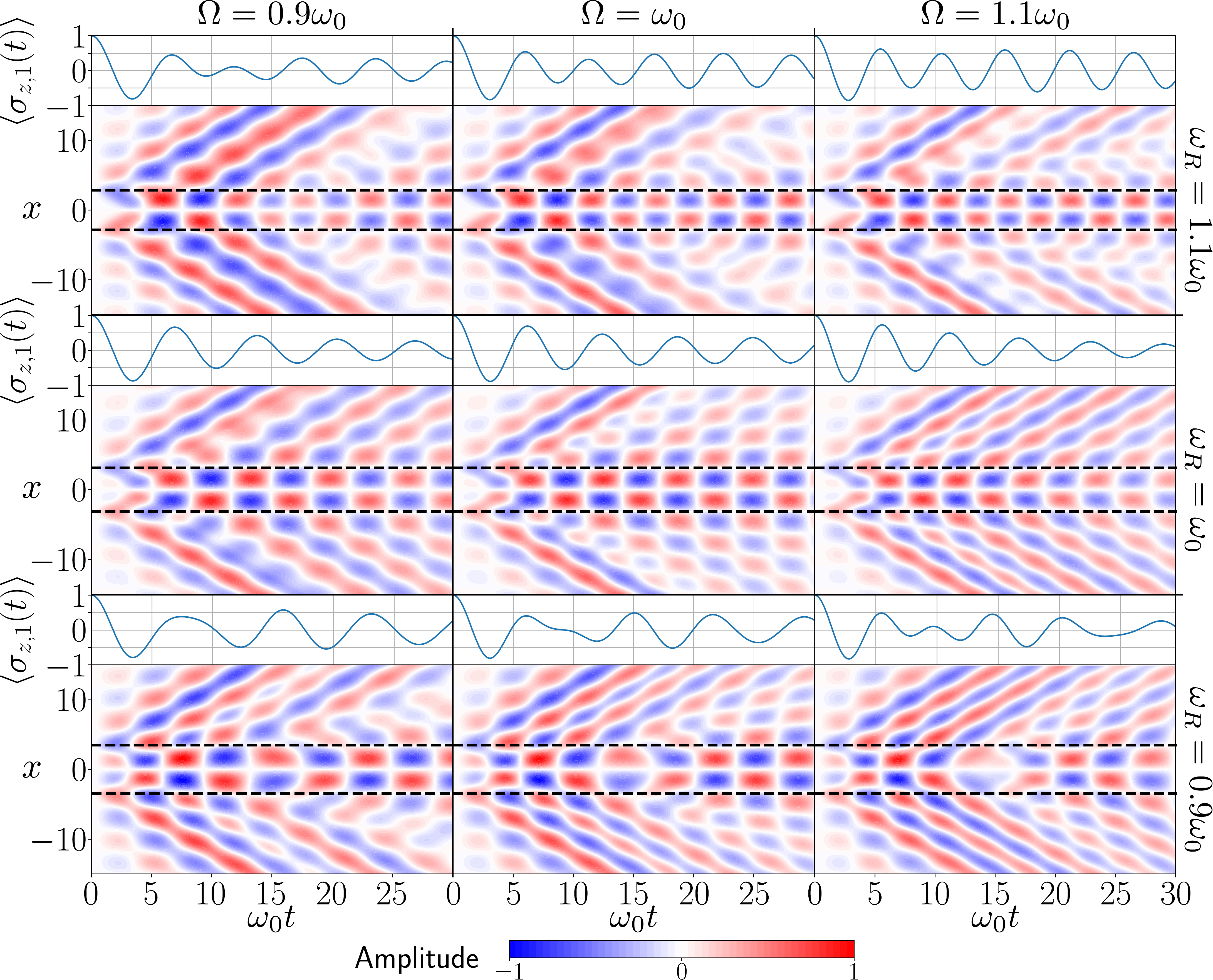}
\caption{\label{fig:CommonEnviron} Spin dynamics (single line plots) and corresponding environment displacement (main panels) at position $x$ over time as calculated with Eq.~\eqref{eq:bathdis} for various separations and dimer frequencies. Displacements are normalized to the maximum displacement across all plots.
The dashed black lines correspond to the positions of the TLSs. For all of the dynamics we have set $T = 0$ and $\epsilon=0$, $\Gamma=0.05\omega_0$ and $\pi\alpha=0.1\omega_0$ for the unmapped spectral density.
}
\end{figure}

In this model there are now three possible resonance conditions which can be met by matching two of: the bare system frequency $\Omega$, the characteristic frequency of the environment $\omega_0$, or the frequency corresponding to the separation between the TLSs  $\omega_R = 2\pi/R$.
In Fig.~\ref{fig:CommonEnvironSpec} we show how choosing $\omega_R=\omega_0$ leads to a complete suppression of the main peak of the bare spectral density, and choosing $\omega_R=(1\pm0.1)\omega_0$ leads to a very asymmetric lineshape. 
In the single line plots of Fig.~\ref{fig:CommonEnviron} we show how these differences in spectral density manifest themselves in the system dynamics. 
We show results for the three TLS spacings $\omega_R=\omega_0, (1\pm0.1)\omega_0$ and for bare system frequencies $\Omega$ tuned to the same three possible values, resulting in nine sets of results.

The system dynamics are complex and difficult to interpret, with multiple oscillation frequencies appearing in the dynamics. These are not only due to the peaked structure of the original spectral density (as occurred for the simplified single spin model) but also because of the spatial correlations. However, we next show that TEMPO can be used not only to extract system dynamics, as discussed in Ref.~\cite{strathearn2017efficient}, but also to find evolution of bath degrees of freedom. This will enable us to gain more insight into how the interplay between the bath and system leads to the complex dynamics shown in Fig.~\ref{fig:CommonEnviron}.
To do this we formally solve the Heisenberg equations of motion~\cite{gardiner1985input,gardiner2004quantum} for the expectation value of e.g.\ the annihilation operator for a particular bath mode, and obtain an expression entirely in terms of the expectation of the system operator which couples to the bath. For our model, Eq.~\eqref{eq:spinbosham}, we have
\begin{align}\label{eq:modexp}
\langle a_k(t) \rangle = -i g_k e^{-i \omega_k t} \int_0^t dt' e^{i \omega_k t'} \langle \sigma_z (t')\rangle.
\end{align}

We thus have access to the dynamics of all the bath modes and can construct the real-space displacement of the bath by creating an appropriately weighted sum of the individual mode displacements. 
The quantity we consider is the displacement of a one-dimensional field~\cite{mahan2013many} given by:
\begin{align}
\Phi(x,t) = \sum_k \frac{1}{\sqrt{\omega_k}}(\langle a_k (t)\rangle e^{ikx} + \langle a^{\dagger}_k(t)\rangle e^{-ikx}).
\end{align}
In the continuum limit this takes the following form:
\begin{multline} \label{eq:bathdis}
\Phi(x,t) = 8 \int_{0}^{\infty}d\omega \left[\sqrt{\frac{J_0(\omega)}{\omega}}\sin\left(\frac{\omega R}{2}\right)\sin(\omega x)\right.\\
\left.\times\int_0^t dt' \sin(\omega (t-t')) \langle \sigma_z (t')\rangle\right]
\end{multline}
where we have used $\omega = |k|$. See Ref.~\cite{Note1} for a derivation of this expression.

We show the real-space displacement of the bath as a function of time and position as a color map beneath the corresponding system dynamics in Fig.~\ref{fig:CommonEnviron}. The black dotted lines indicate the positions of the two TLSs.
In the fully resonant case (the centre plot) we see that most of the bath excitation gets trapped between the two TLSs which simply show oscillations which decay away at a very slow rate.
We also see slow decay in the other two plots where $\Omega=\omega_R$; this is because $J(\Omega)=0$ in these cases and hence a simple Markovian treatment would predict no decay at all.  

When at least one of the frequencies is detuned from this condition we see a more pronounced propagation of bath excitations away from the system as the trapping effect is reduced. In these cases the overall decay rate of the TLS is gradually enhanced as the detuning is increased, since now the value of $J(\Omega)$ is larger.
The detuning can also introduce a beat frequency into the bath dynamics, most clearly seen in the bath population trapped between the two TLSs, and most pronounced in the most detuned cases, i.e.\ in the bottom right and top left panels of Fig.~\ref{fig:CommonEnviron}. This beating gives rise to distinctive revivals in the TLS dynamics.

At very short times all of the dynamics are very similar. On timescales $t<\omega_R^{-1}$ each TLS only senses its own local environment and hence behaves as if it were interacting with bosons described by the unmapped spectral density. 
However as soon as the influence of the other TLS is felt the dynamics become radically different to those predicted by an independent environment model. 
These dynamics are highly sensitive to the relative detunings of the three frequencies described above, becoming significantly different even for the very small detunings studied here.

In conclusion, we have shown that TEMPO can provide accurate simulations of quantum systems in structured environments. We first validated the technique by simulating the quantum dynamics of a simple model that can be solved accurately using the RC approach, and comparing the results. We then presented TEMPO simulations of the more complex dynamics that result from the interplay between a highly structured spectral function and spatial correlations between different parts of the system.
TEMPO is not limited to specific forms for the spectral density and so provides a versatile method for simulating systems coupled to an environment with arbitrary structure. 
In addition, we have shown that it is possible to find the full dynamics of the environment directly from the system dynamics.
This enabled us to explore how it is possible to tune the separation between the spins such that one can effectively remove the coupling to the dominant environmental mode and how the resultant dynamics are highly sensitive to slight variations in parameters around this point. 
Such exact environmental tracking can help to explain the behavior of open quantum systems in general, and so aid in the design of future quantum devices.
This greater insight into the dynamics of open quantum systems could also lead to the development of new approximation schemes and to the self-consistent verification of others.

\acknowledgments{DG and DK acknowledge  studentship funding from EPSRC under grant no.\ EP/L015110/1. AS acknowledges a studentship from EPSRC under grant no.\ EP/L505079/1. J.I.-S. acknowledges support from the Royal Commission for the Exhibition of 1851. AN acknowledges funding from EPSRC under grant no.\ EP/N008154/1. PK acknowledges support from an ESQ fellowship of the Austrian Academy of Sciences (\"OAW). }

%

\widetext
\clearpage
\begin{center}
    \textbf{\large Supplementary Material for: Exact Quantum Dynamics in Structured Environments}
\end{center}
\setcounter{equation}{0}
\setcounter{figure}{0}
\setcounter{table}{0}
\setcounter{page}{1}
\makeatletter
\renewcommand{\theequation}{S\arabic{equation}}
\renewcommand{\thefigure}{S\arabic{figure}}
\renewcommand{\bibnumfmt}[1]{[S#1]}
\renewcommand{\citenumfont}[1]{S#1}


\author{Dominic Gribben}
 \affiliation{SUPA, School of Physics and Astronomy, University of St Andrews,
St Andrews, KY16 9SS, United Kingdom}
\author{Aidan Strathearn}
 \affiliation{SUPA, School of Physics and Astronomy, University of St Andrews,
St Andrews, KY16 9SS, United Kingdom}
 \affiliation{School of Mathematics and Physics, The University of Queensland, St Lucia,
Queensland 4072, Australia}
\author{Jake Iles-Smith}
 \affiliation{Department of Physics and Astronomy,  University of Sheffield, Sheffield,  S3  7RH,  United  Kingdom}
\author{Dainius~Kilda}
 \affiliation{SUPA, School of Physics and Astronomy, University of St Andrews,
St Andrews, KY16 9SS, United Kingdom}
\author{Ahsan Nazir}
 \affiliation{School  of  Physics  and  Astronomy,  The  University  of  Manchester,  Oxford  Road,  Manchester  M13  9PL,  United Kingdom}
\author{Brendon W. Lovett}
 \affiliation{SUPA, School of Physics and Astronomy, University of St Andrews,
St Andrews, KY16 9SS, United Kingdom}
\author{Peter Kirton}
\affiliation{Vienna Center for Quantum Science and Technology, Atominstitut, TU Wien, 1040 Vienna, Austria}

\section{Reaction Coordinate Mapping}

In this section we outline how the Reaction Coordinate (RC) mapping is implemented to obtain a time-local master equation for the two-level system-RC (TLS-RC) reduced density matrix.
The RC is the coordinate corresponding to the direction the bath equilibrium position is displaced when the system transitions between $\sigma_z$ eigenstates. It corresponds to a superposition of normal modes of the bath, and is defined by the following mapping:
\begin{align}\label{HSB}
H_{I} = \sigma_z \sum_k g_k (a_k^\dagger+a_k) \rightarrow \lambda \sigma_z ( c^\dagger + c).
\end{align}
Here, $c^\dag$ ($c$) create (destroy) an excitation in the reaction coordinate mode.
The form of the transformation results in coupling between the RC and the residual modes, $b_k$.
The spin-boson Hamiltonian is then mapped to the following form:
\begin{align}
&H_{RC} = H_S + H_I + H_B+H_C \\
&H_S = \Omega \sigma_x + \epsilon \sigma_z +  \lambda \sigma_z ( c^\dagger + c)+\Omega_{RC} c^{\dagger} c \\
&H_I = (c^\dagger + c) \sum_k d_k (b_k^\dagger + b_k) \\
&H_B = \sum_k \nu_k b_k^\dagger b_k \\
&H_C = (c^\dagger + c)^2 \sum_k \frac{c_k^2}{\omega_k}
\end{align}
where the last term exists to counter a renormalization of the RC's potential that arises from the interaction term. Here the TLS-RC coupling is given by $\lambda^2=\sum_k g_k^2$ and this form ensures that the canonical bosonic commutation relation is satisfied. This coupling leads to an RC frequency of $\Omega_{RC} = \lambda^{-2}\sum_k \omega_k g_k^2$. The residual environment is characterised by a new spectral density $J_{RC}(\omega) = \sum_k |d_k|^2 \delta (\omega-\omega_k)$ which can be found by replacing the TLS in both the mapped and un-mapped Hamiltonians with a classical coordinate~\cite{garg1985effect_supp}. The spectral density does not contain any information about the system itself and therefore can be found, in both cases, through the classical equations of motion of this coordinate~\cite{leggett1984quantum_supp, iles2014environmental_supp}. The under-damped spectral density considered in the main text leads to $\Omega_{RC} = \omega_0$, $\lambda = \sqrt{\pi \alpha_{\text{UD}} \omega_0/2}$ and $J_{RC} = \gamma \omega$ where $\gamma = \Gamma/2\pi\omega_0$.

Once this mapping has been performed we follow the procedure outlined in Ref.~\cite{iles2014environmental_supp} applying both Born and Markov approximations to arrive at the Schr\"odinger picture master equation for the reduced TLS-RC density operator, $\rho(t)$:
\begin{multline}\label{eq:schme}
\frac{\partial \rho}{\partial t} = -i [H_S,\rho(t)]
- \gamma \int_0^{\infty}d\tau \int_0^{\infty} d\omega \omega \cos (\omega \tau) \coth \left(\frac{\omega}{2T}\right) [\hat{C},[\hat{C}(-\tau),\rho(t)]]\\
 - \gamma \int_0^{\infty}d\tau \int_0^{\infty} d\omega \omega \cos (\omega \tau) [\hat{C},\lbrace[\hat{C}(-\tau),H_S],\rho(t)\rbrace]
\end{multline}
where we have defined $\hat{C}=c^{\dagger}+c$ and $\hat{C}(t)$ is the same operator in the interaction picture at time $t$. Truncating the Hilbert space of the RC down to $n$ basis states, i.e.\ permitting a maximum of only $n$ excitations, allows us to numerically diagonalize $H_S$. This gives us a set of basis states $\ket{\phi_j}$ which satisfy $H_S\ket{\phi_j} = \phi_j \ket{\phi_j}$  allowing us to express the interaction picture operators as:
\begin{align}
\hat{C}(t) = \sum_{j,k=1}^{2n} C_{jk}e^{i\omega_{jk}t}\dyad{\phi_j}{\phi_k}
\end{align}
where $C_{jk} = \bra{\phi_j}\hat{C}\ket{\phi_k}$ and $\omega_{jk}=\phi_j-\phi_k$.
 We can then re-write Eq.~\eqref{eq:schme} as:
\begin{align}
\frac{\partial \rho}{\partial t} = -i [H_S,\rho(t)]-[\hat{C},[\hat{\chi},\rho(t)]]+[\hat{C},\lbrace\hat{\Xi},\rho(t)\rbrace]
\end{align}
with rate operators:
\begin{align}
\hat{\chi} &= \frac{\pi}{2} \sum_{jk} J_{RC} (\xi_{jk})\coth (\frac{\xi_{jk}}{2T})C_{jk}\dyad{\phi_j}{\phi_k}
\\
\hat{\Xi} &= \frac{\pi}{2} \sum_{jk} J_{RC} (\xi_{jk})C_{jk}\dyad{\phi_j}{\phi_k}.
\end{align}
The resulting master equation can then be solved using standard open systems approaches for Markovian systems~\cite{breuer2002theory_supp}.

\section{Time-Evolving Matrix Product Operators}
In this section we outline the implementation of the Time-Evolving Matrix Product Operators (TEMPO) algorithm. For full details see~\cite{strathearn2017efficient_supp}.

The quasi-adiabatic path integral (QUAPI) method~\cite{makri1995tensor_supp} is used to calculate the non-Markovian evolution of the system up to time $t_N = N\Delta t$ by summing over all possible paths the system has taken to that point from initial time $t_0$. In practice this can be done through an iterative tensor propagation routine where the object propagated is the augmented density tensor (ADT). The ADT is grown up to time $t_N$ from an initial physical density matrix $\rho_{i_1} (t_1)$ through iterative application of tensors:
\begin{equation}
A^{j_{N}\dots j_1}(t_N) = \prod_{n=2}^N B^{j_{n},j_{n-1},...,j_1}_{\;\;\;\;\;i_{n-1},...,i_1} \rho_{i_1}(t_1).
\end{equation} 
The reduced system density matrix at time $t_N$ is then given by $\rho_{j_N}(t_N) = \sum_{j_1\dots j_{N-1}} A^{j_{N}\dots j_1}(t_N)$. For a harmonic bath linearly coupled to the system, the $B$ tensor is composed of influences across all pairs of time points from $t_1$ to $t_N$, as well as time-local components due to the coherent Hamiltonian evolution. The discretisation time-step $\Delta t$ needs to both resolve the features of the correlation function and ensure the Trotter errors are negligible. Without further considerations the exponential growth of the ADT only allows memory lengths of order $\sim20\Delta t$ to be simulated. This restricts QUAPI to baths with relatively short-lived and/or slowly varying auto-correlation functions.

In TEMPO the ADT is built in the form of a matrix product state (MPS)~\cite{orus2014practical_supp,schollwock2011density_supp} and a singular value decomposition (SVD) sweep is carried out at each time-step. The $B$ tensors are composed of a product of time-local operators and keeping these components separate leads to the natural formation of an MPS representation for the $A$ tensor given by:
\begin{equation}
A^{j_{N}\dots j_1}(t_N) = \sum_{\alpha_1\dots \alpha_N} [a^{j_N}]_{\alpha_N}[a^{j_{N-1}}]_{\alpha_N,\alpha_{N-1}}\dots[a^{j_1}]_{\alpha_1}.
\end{equation}
where the tensors at the two ends of the chain are rank two, and those in between are rank three. In this notation the superscripts are the `physical' index which connects to the propagation tensor while the subscripts link adjoining $a$ tensors.
At each step of the propagation an SVD is performed on each $a$ and singular values below a fixed precision $\chi$ (relative to the largest) are discarded. This leads to a reduced dimension of the internal indices $\alpha$ and a significant improvement on the exponential scaling present in QUAPI. The truncation acts to remove the least important internal degrees of freedom that are not needed for achieving converged dynamics. This proceedure has been shown to reduce the scaling of the required  computational resources to polynomial with memory length for smooth spectral densities~\cite{strathearn2017efficient_supp}.

\begin{table}[h]
\begin{center}
\begin{tabular}{|m{3cm}|m{3cm}|m{3cm}|}

\hline
\hfil Result & \hfil Time-step, $\Delta t$ & \hfil Precision, $\chi$ \\
\hline
\hfil Fig. 2 & \hfil 0.2/$\Omega$ & \hfil $10^{-8}$ \\
\hline
\hfil Fig. 4 & \hfil 0.1/$\omega_0$ & \hfil $10^{-7}$ \\
\hline
\end{tabular}
\end{center}
\caption{Convergence parameters used for figures in the main text.}
\label{tab:con}
\end{table}

In practice there are two convergence parameters associated with TEMPO which must be adjusted to produce the exact results in the main text:
\begin{itemize}
\item The size of the discretization time-step $\Delta t$, decreased until convergence.
\item The singular value precision $\chi$, increased until convergence.  
\end{itemize}

Further to these, with TEMPO a memory cut-off is also possible such that the growth of the ADT is stopped once it covers enough of the system's history to capture all relevant memory effects. For the results presented in the main text, however, the long correlation times associated with the structured environments make this kind of memory cut-off impossible and as such none were used for generating any of the results. Table~\ref{tab:con} gives the convergence parameters used for each of the results.

\section{Bath Dynamics}
In this section we show that it is possible to calculate bath operator expectation values in terms of system dynamics and specifically present the derivation of Eq.~(9) in the main text.

Consider general interaction and bath Hamiltonians given by:
\begin{equation}
H_I(t) + H_B = \hat{A}(t) \sum_k  (g_k a_k^{\dagger}+g_k^* a_k) + \sum_k \omega_k a_k^{\dagger} a_k
\end{equation}
for a general system operator $\hat{A(t)}$ in the system interaction picture. The time evolution of the ladder operator $a_k(t)$ can be found from the Heisenberg equation of motion:
\begin{align}
\frac{d}{dt}a_k(t) &=  i[H_I(t) + H_B,a_k(t)]\\
&= -ig_k \hat{A}(t) - i \omega_k a_k(t)\\
e^{i \omega_k t}\frac{d}{dt}a_k(t) + i\omega_k e^{i\omega_k t} a_k(t) &= -i g_k e^{i\omega_k t} \hat{A}(t)  \\
\frac{d}{dt}\left[e^{i\omega_k t}a_k(t)\right] &= -i g_k e^{i\omega_k t} \hat{A}(t).
\end{align} 
Formally integrating this and taking the expectation with respect to the total density matrix $\rho = \rho_S \otimes \rho_B$ gives:
\begin{align}
\langle a_k(t) \rangle &= -i g_k e^{-i \omega_k t} \int_0^t dt' e^{i \omega_k t'} \langle \hat{A} (t')\rangle, \\
\langle a_k^{\dagger}(t) \rangle &= i g_k^* e^{i \omega_k t} \int_0^t dt' e^{-i \omega_k t'} \langle \hat{A} (t')\rangle.
\end{align}
Thus the dynamics of any bath mode can be found from knowing the exact system dynamics up to the time $t$. Calculations of higher order moments are much more involved requiring multiple numerical integrals to be performed and so for now we only consider quantities that can be constructed from the expressions above. The quantity analysed in the main text is given by:
\begin{align}
\Phi(x,t) = \sum_k \frac{1}{\sqrt{\omega_k}} \left[ \langle a_k(t)\rangle + \langle a_{-k}^{\dagger}(t)\rangle \right]e^{i k x}.
\end{align}
Now for the $\hat{A} = \sigma_{z}$ coupling from Eq.~(5) in the main text we have $\tilde{g}_{-k}^* = \tilde{g}_k$ and $\omega_k = |k| = \omega_{-k}$ such that:
\begin{align}
\Phi(x,t) &= \sum_k \frac{i\tilde{g}_k}{\sqrt{\omega_k}} \left[-  e^{-i \omega_k t} \int_0^t dt' e^{i \omega_k t'} \langle \sigma_z (t')\rangle + e^{i \omega_k t} \int_0^t dt' e^{-i \omega_k t'} \langle \sigma_z (t')\rangle \right] e^{i k x} \\
 &= -\sum_k \frac{2\tilde{g}_k}{\sqrt{\omega_k}} e^{ikx} \int_0^t dt' \sin[\omega_k(t-t')] \langle \sigma_z (t') \rangle \\
 &= -4i \int_{-\infty}^\infty dk \frac{g_k}{\sqrt{\omega_k}}\sin\left(\frac{kR}{2}\right) \left[ \cos(kx)+i\sin(kx) \right] \int_0^t dt' \sin[\omega_k(t-t')] \langle \sigma_z (t') \rangle \\
 &= 8 \int_0^\infty d\omega \sqrt{\frac{J_0(\omega)}{\omega}} \sin\left(\frac{\omega R}{2}\right) \sin(\omega x) \int_0^t dt' \sin[\omega (t-t')] \langle \sigma_z (t') \rangle 
\end{align}
where between the last two lines we have discarded the odd component of the integrand and since everything remaining is even we can then halve the integration domain.

\end{document}